\documentclass[11pt,a4paper]{scrartcl}
\usepackage[numbers]{natbib}

\usepackage{booktabs}

\usepackage{color}
\usepackage{latexsym}              
\usepackage{amsmath}
\usepackage{amssymb}               
\usepackage{amsfonts}              
\usepackage{amsthm}                
\usepackage{mathtools}
\usepackage{multirow}
\usepackage{tikz}                  
\usetikzlibrary{arrows,positioning,shapes}
\usepackage{tcolorbox}
\usepackage{pifont}

\usepackage[english]{babel} 

\RequirePackage[%
  pdfstartview=FitH,%
  breaklinks=true,%
  bookmarks=true,%
  colorlinks=true,%
  linkcolor= blue,
  anchorcolor=blue,%
  citecolor=orange,
  filecolor=blue,%
  menucolor=blue,%
  urlcolor=blue%
  ]{hyperref}
  
  \AtBeginDocument{%
  \hypersetup{%
    pdfauthor={M Perez-Ortiz},%
    pdftitle={Title - compilation : \today}%
    }
    }

\usepackage{fancyhdr}
\begin{document}
\title{Semantic TrueLearn: Using Semantic Knowledge Graphs in Recommendation Systems}

\author{
\small{Sahan Bulathwela,
    Mar\'ia P\'erez-Ortiz,
Emine Yilmaz,
    John Shawe-Taylor
}
\\[2ex] 
\small{
    Centre for Artificial Intelligence, University College London (UK)
}
}

\date{ }

\maketitle

\begin{abstract}
In informational recommenders, many challenges arise from the need to handle the semantic and hierarchical structure between knowledge areas. This work aims to advance towards building a state-aware educational recommendation system that incorporates semantic relatedness between knowledge topics,  propagating latent information across semantically related topics. We introduce a novel learner model that exploits this semantic relatedness between knowledge components in learning resources using the Wikipedia link graph, with the aim to better predict learner engagement and latent knowledge in a lifelong learning scenario.
In this sense, \emph{Semantic TrueLearn} builds a humanly intuitive knowledge representation
while leveraging Bayesian machine learning to improve the predictive performance of educational engagement. Our experiments with a large dataset demonstrate that this new semantic version of TrueLearn algorithm achieves statistically significant improvements in terms of predictive performance with a simple extension that adds semantic awareness to the model.

\end{abstract}

\section{Introduction} \label{searea c_intro}

Developing artificial intelligence systems that, mildly at least, understand the structure of knowledge is
foundational to
building an effective recommendation system for education \cite{bauman2018recommending,goal_based_edrec}, as well as for many other applications  \cite{lewis2020retrieval,yano2016taking} related to knowledge management and tracing.

Through this work, we propose \emph{Semantic} TrueLearn, a novel and transparent learner model that incorporates automatic entity linking and Wikipedia (a publicly available, humanly-intuitive, domain-agnostic and ever-evolving) knowledge graph, as a first step towards building an educational recommender that automatically labels materials and embeds the structure of universal knowledge using Wikipedia. Semantic TrueLearn is a probabilistic graphical model that maintains a symbolic representation of learners' knowledge that allows explanations, rationalisations and scrutinisation.
We i) propose a novel approach for modelling semantic relatedness, ii) propose a novel sub-symbolic Bayesian learner model, iii) identify several research questions relating to validating the improvement of this proposed model and iv) evaluate the performance of the proposed model in a large dataset of learners engagement with educational resources.





\section{Related work} \label{sec:lit_survey}
 

Knowledge Tracing (KT) \cite{Yudelson13} is one of the most popular methods for user modelling in educational recommendation contexts. Incorporation of Semantic Relatedness in KT systems has been attempted before in prerequisite modelling \cite{prerequisites,chen2018prerequisite}, exercise similarity \cite{pandey2020rkt,huang2019ekt,gkt} and  other tasks \cite{bauman2018recommending,thaker2020recommending}. However, KT often relies on expert labelling of the \emph{Knowledge Components} (KCs) \cite{assistments_data} (sometimes also for knowledge hierarchies \cite{bauman2018recommending}), which is not scalable to large-scale lifelong learning applications in practice. 
\emph{Wikification}, a form of entity linking \cite{wikifier} that expands this idea, has shown substantial progress and great promise for automatically capturing the KCs covered in an educational resource. Educational Recommenders using Wikification, such as TrueLearn \cite{truelearn}, has shown promise in building probabilistic graphical models, although one of the assumptions of this model is that Wikipedia concepts are independent and thus unrelated.

Using Wikpedia as an ontology or knowledge graph to understand documents is not a new idea. While Wikipedia itself has been used as an ontology using page link and category links to describe "relates to" and "is type of" relationships respectively \cite{wiki_onto,0cb3e7c8791240a7962ed84af3dc2f06}, other works have pushed further and used the wealth of information in Wikipedia to build downstream knowledge bases and ontologies \cite{auer2007dbpedia}, as well as ontology-driven information retrieval systems \cite{grefenstette2015transforming}. From the early days of Wikipedia, exploiting different aspects of its contents (such as text, link structure, etc.) to model \emph{Semantic Relatedness} (SR) that represents "relates to" links has been attempted. These SR metrics have evolved over time into recent proposals that are diverse and sophisticated metrics highly predictive of concept relatedness \cite{ponza_semantic_relate}. However, the utility in these proposals with graphical models is yet to be explored and is addressed thorugh this work.

\paragraph{Representations from Graphs} 
The core technical contribution of this work is proposing a method to infer the latent value of an unobserved skill parameter using observed ones via information sharing based on a semantic relatedness graph. 
Several works have proposed novel ways to use a relationship graph to recover a latent representation for an unknown node using a set of known nodes. 
Recent Graph Convolutional Neural Networks \cite{kipf2017semi} infer hidden node embeddings ($H^{\ell + 1}$) by weighted averaging the embeddings of its neighbours using adjacency matrix $A$ and diagonal $D$ ($H^{\ell + 1} = D^{-\frac{1}{2}}AD^{-\frac{1}{2}}H^\ell W^\ell$).
The popularising attention mechanism \cite{attention} uses alignment, which is a weighted sum of embeddings.
This work puts the foundations to applying these approaches in a educational recommender using 1) a probabilistic graphical model and 2) SR values extracted from Wikipedia.

\section{Methodology} \label{sec:method}

\paragraph{Problem Formulation}Consider a learning environment in which a learner $\ell$ interacts with a set of educational resources $S_\ell \subset \{r_1, \ldots, r_Q\}$ over a period of $T= (1,\ldots,t)$ time steps, $Q$ being the total of resources in the system. Each resource  $r_i$  is characterised by the top KCs or topics covered $K_{r_i} \subset \{1, \ldots, N \}$ ($N$ being the total of KCs considered by the system) and the depth of coverage $d_{r_i}$  of those. 
We represent learner knowledge at time $t$ as a multivariate Gaussian distribution $\boldsymbol{\theta}^t_\ell \sim \mathcal{N}(\boldsymbol{\mu}^t_\ell, {\Sigma}^t_\ell)$, $\boldsymbol{\mu}^t_\ell \in \mathbb{R}^Q$ being the mean of knowledge and $\Sigma^t_\ell$ the covariance matrix. TrueLearn assumed that $\Sigma$ is a diagonal covariance matrix in all cases and thus knowledge topics are completely independent from each other. The work in this paper builds towards considering a full covariance matrix, assuming that $\rho_{ij}$ (estimated semantic relatedness) is a proxy for $\Sigma_{ij}$ for topics $i$ and $j$ when $i \neq j$.

The key idea behind TrueLearn \cite{truelearn} is to model the probability of engagement $e_{\ell, r_i}^{t} \in \{ 1, -1\}$ between learner $\ell$ and resource $r_i$ at time $t$ as a function of the learner skills/ knowledge $\boldsymbol{\theta^t}_\ell$ and 
resource representation $d_{r_i}$ for the top KCs covered $K_{r_i}$.
When a new learner joins the recommender system, TrueLearn sets $\mu_\ell^0 = 0$ and $\Sigma_{ii} = \beta$, where $\beta$ is a hyperparameter of the system, and $\Sigma_{ij} = 0, \; i \neq j$. Then, when the learner consumes an educational video fragment, TrueLearn updates the learner model/skills accordingly. Every skill that is not updated is set to the value from the last step, meaning at time $t$ there might be many unobserved skills, specially given the amount of topics considered by the system (equal to the number of Wikipedia pages). 
Thus, TrueLearn assumes that the skill for topics in $K_{r_i}$ can only be obtained through those topics and not semantically related ones. 

\subsection{Semantic TrueLearn}\label{topic:sr_truelearn}

\begin{figure}[ht!]
\small
\begin{center}
\centerline{\includegraphics[width=0.5\columnwidth]{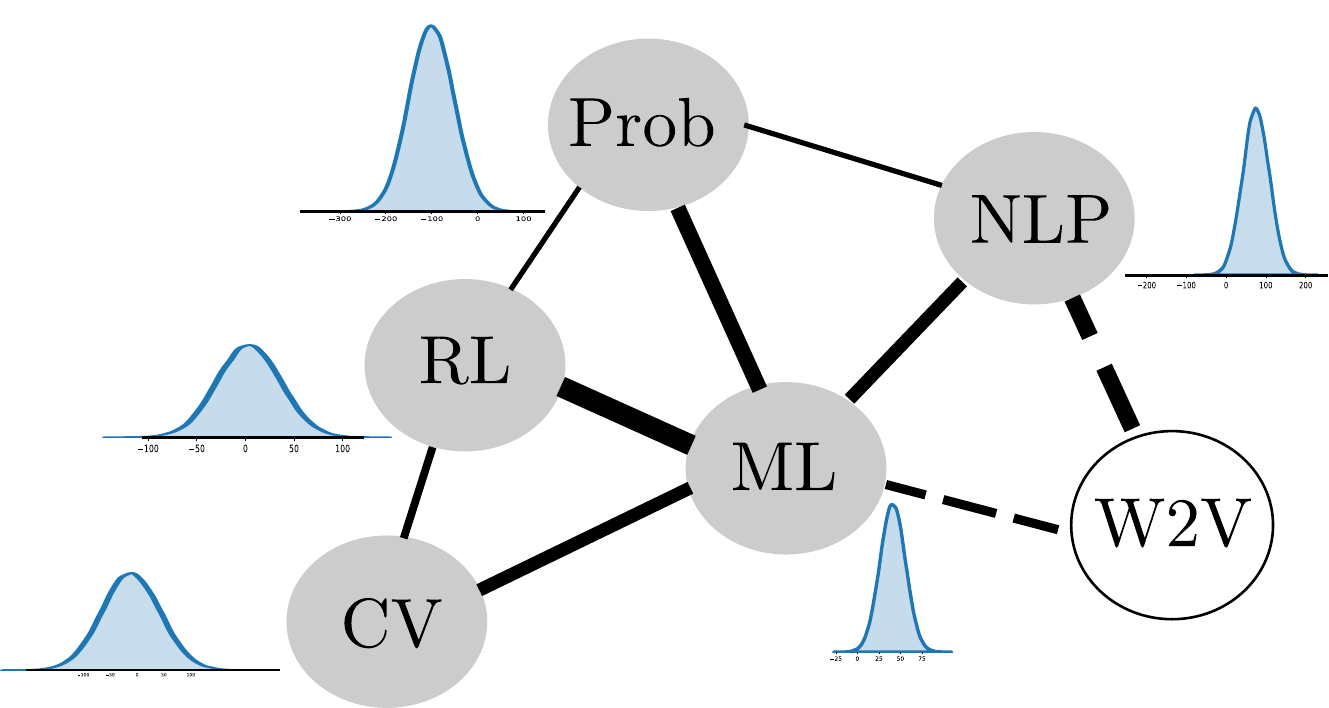}}
\caption{Inferring the knowledge for unseen topic (white circle) based on semantically related and seen ones (grey circles) by transferring knowledge (dotted lines).
Topics are: ML (Machine Learning), RL (Reinforcement Learning), Prob (Probability), CV (Computer Vision), NLP (Natural Language Processing) and W2V (Word2Vec).}
\label{fig:semantic_rel_problem}
\end{center}
\end{figure}

Extending the TrueLearn model \cite{truelearn,trueeducation}, the proposed Semantic TrueLearn, is a learner model that infers the knowledge state of learners in an online fashion. Semantic TrueLearn exploits its current knowledge of observed concepts and their SR to a novel concept to make a better prior estimation as per Figure \ref{fig:semantic_rel_problem}. 

\paragraph{Incorporating Semantic Relatedness Between Topics} 
The main assumption for Semantic TrueLearn, the proposal, is that knowledge can be shared across semantically related topics. By taking inspiration from graph convolutions \cite{kipf2017semi}, we formally assume a relationship as per equation \ref{eq:semantic_rel}.

\begin{equation} \label{eq:semantic_rel}\centering
    \theta_{\ell, i}^t = \frac{1}{|\Omega_{\ell,i}|} \sum_{j \in \Omega_{\ell, i}} \gamma_{ij} \cdot \theta^t_{\ell,j},\text{\ where\ } \theta^t_{\ell,i} \sim \mathcal{N}(\mu, \sigma^2)
\end{equation}
where $\Omega_i$ represents the set of topics used to infer the representation of topic $i$ (e.g. most semantically related seen topics), where $i\neq j$. The mixing factors $\gamma_{ij}$ can be set to semantic relatedness $\rho_{ij}$ 
or to a factor of the standard error of topic $j$ (meaning most observed topics are used). In the TrueLearn \cite{truelearn} model, which we extend, $\theta$ is a Gaussian variable. We use equation \ref{eq:gauss} to calculate unknown parameter $\theta_{\ell, i}^t$. 

\begin{equation}\label{eq:gauss}\centering
    \hat{\theta}^t_{\ell,i} \sim \mathcal{N} \left (
    \sum_{j \in \Omega_i} \frac{1}{|\Omega_i|} \cdot \rho_{ij} \cdot \mu^t_{\ell,j}, 
    \sum_{j \in \Omega_i} \left ( \frac{1}{|\Omega_i|} \cdot \rho_{ij} \right )^2 
    \cdot \left ( \sigma^t_{\ell,i} \right )^2 \right )
\end{equation}

\paragraph{Semantic Relatedness Metric (SR Metric)}
As mentioned in section \ref{sec:lit_survey}, different measures of SR for Wikipedia concepts exist \cite{ponza_semantic_relate}. We empirically evaluate if the predictive performance of an educational recommender can be improved by incorporating 7 different SR Metrics to substitute $\rho_{ij}$ in equation \ref{eq:gauss}. We devise Milne and Witten (\texttt{M\&W}), Entity Embeddings (\texttt{W2V}), Point-wise Mutual Information (\texttt{PMI}), Language Model based (\texttt{LM}), Jaccard Similarity (\texttt{Jaccard}),  Conditional Probability (\texttt{CP}) and  Barabasi and Albert  (\texttt{BA}) SR Metrics, where SR values are pre-computed and publicly available \cite{wat_piccinno}. 

\section{Experiments and Results} \label{sec:results}

We have four main research questions we verify in this preliminary experiment. 
\begin{itemize}
    \item \textbf{RQ1}: Which SR Metric is most suitable?
    \item \textbf{RQ2}: How many related topics should be considered?
    \item \textbf{RQ3}: Does Semantic TrueLearn outperform TrueLearn? 
    \item \textbf{RQ4}: Does semantic information contribute to the gains? In which cases?
\end{itemize}

\subsection{Data}
We use the PEEK dataset \cite{peek_orsum}, a dataset of more than 20,000 learners consuming video lectures in VideoLectures.Net\footnote{\url{www.videolectures.net}} platform. The dataset provides information about how different users consumed fragments of videos over time \cite{x5learn}. This dataset uses entity linking \cite{wikifier} to associate most related Wikipedia concepts to documents. TagMe WAT API \cite{wat_api} \footnote{\url{https://sobigdata.d4science.org/web/tagme/wat-api}} provides the required SR annotations. 

To keep the computational complexity lower, a smaller dataset of 20 most active users is used for RQ1 and RQ2. However, the full dataset of 20,000 users is used to validate RQ3 and RQ4, which are our primary research questions.


\subsection{Evaluation} 
A sequential prediction design where engagement at time $t$ is predicted using events $1$ to $t-1$. A training set of 70\% of the learners is used for hyperparameter tuning and the remainder is used for testing and reporting. Being a binary classification task, precision, recall and F1-measure are evaluated whereas F1-measure is used for overall model selection \cite{peek_orsum}. The evaluation metrics are computed for each learner separately and the weighted average of the scores based on the number of learner's events is reported. To verify statistical significance of the improvement in RQ3, we use a learner-wise one-tailed paired t-test.

\subsubsection{Impact of Semantic Relatedness (RQ4)}

We use the topics encountered in user sessions to build a topic relatedness graph and extract a few attributes linked to graph connectedness for each user. Spearman's Rank Order Correlation Coefficient (SROCC) statistic is then used to evaluate the correlation between the extracted features and the predictive performance. User's  \emph{number of events}, \emph{number of unique topics}, \emph{topic sparsity rate}
\cite{truelearn}, \emph{positive label rate}, \emph{Avg. Connectedness}, i.e. average of the degree distribution of the topics, and \emph{Min. Cut Set Size}, i.e. the minimum number of topics that need to be removed to break the graph into more sub-graphs, are analysed. 
The correlation with the recall score is investigated as the improvement in recall attributes to the performance gains of the proposed model (see Table \ref{tab:results}).
To validate if semantic relatedness is specifically influential in earlier parts of the user session, we plot the mean recall score of all users at event $n$, for different number of events ($n$).

\subsection{Results}

\begin{figure}[b!]
\small
\begin{center}
\centerline{\includegraphics[width=\columnwidth]{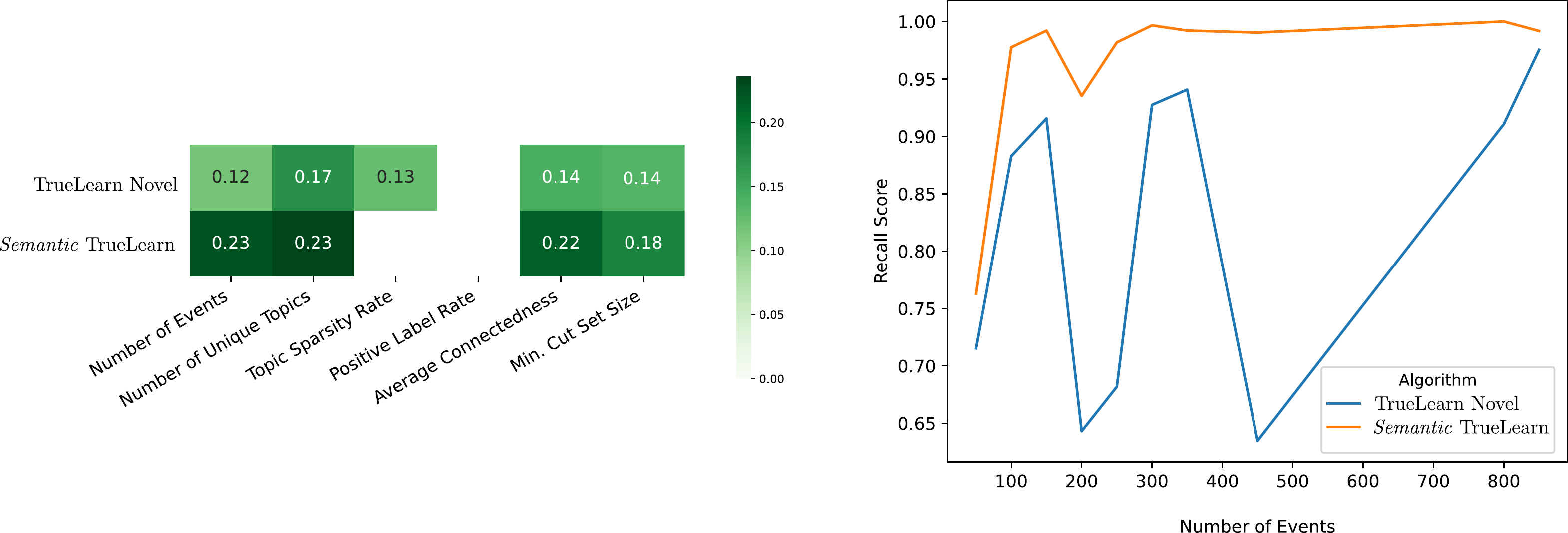}}
\caption{(Left) Relationship between different behavioural characteristics of user profiles and model recall performance presented using SROCC. The numbers and the intensity of each cell corresponds to the Spearman R coefficient where significant correlation is present ($p<0.01$). Empty cells represent lack of significant correlation.(Right) The average recall performance of the two models for the learner population at different number of events.}
\label{fig:performance}
\end{center}
\end{figure}

We run experiments to answer the research questions outlined above. To identify the most suitable SR metric (RQ1), we evaluate Semantic TrueLearn model using 7 SR Metrics proposed in section \ref{topic:sr_truelearn}. The results are outlined in Table \ref{tab:results}.
To understand the Effect of $\Omega_{i}$, the Number of Semantically Related Topics (RQ2), we use the identified SR Metric to experiment with different numbers of semantically related topics. The results of this experiment are reported in Table \ref{tab:sr_topics}. Finally, we use the full PEEK dataset to validate if the use of SR data improves baseline TrueLearn model (RQ3). The results obtained in this experiment are presented in Table \ref{tab:st_model}. Figure \ref{fig:performance} presents the results obtained in investigating the impact of semantic relatedness (RQ4) where (left) the correlation investigation between topic connectivity of users and recall score, and (right) the performance of the model based on different number of events is reported. 


\begin{table}[t!] \small
    \caption{Predictive performance of adding Semantic Relatedness (SR) to TrueLearn Novel algorithm. The different configurations (SR Metric) of Semantic TrueLearn Novel algorithm (our proposal) are evaluated using Precision (Prec.), Recall (Rec.) and F1 Score (F1). The most performant value and the next best value are highlighted in \textbf{bold} and \emph{italic} faces respectively. The Semantic TrueLearn algorithms that outperform baseline model in terms of F1 score are \underline{underlined}.}
    \label{tab:results}
    \centering
	\begin{tabular}{c c c c c}
	\hline
	Algorithm & SR Metric & Prec. & Rec. & F1 \\
	\hline
	TrueLearn & & & & \\
	Novel & - & 0.7667 & 0.9476 & 0.8348 \\
	\hline
	&\texttt{M\&W} & 0.7701 & 0.9469 & \underline{\textit{0.8364}} \\
	&\texttt{W2V} & 0.\textbf{7714} & 0.9467 & \underline{\textbf{0.8370}} \\
	\emph{Semantic}&\texttt{PMI} & 0.7682 & \textit{0.9480} & \underline{0.8355} \\
	TrueLearn&\texttt{LM} & 0.7605 & \textbf{0.9507} & 0.8322 \\
	Novel&\texttt{Jaccard} & 0.7605 & \textbf{0.9507} & 0.8322 \\
	&\texttt{CP} & 0.7621 & \textbf{0.9507} & 0.8330 \\
	&\texttt{BA} & \textit{0.7704} & 0.9469 & \underline{\textit{0.8364}} \\
	\hline
	\end{tabular}
\end{table}


\begin{table}[ht!] 
\caption{The performance of Semantic TrueLearn model with \texttt{W2V} SR metric is reported in terms of Precision (Prec.), Recall (Rec.) and F1 Score (F1). The performance of the model is reported when different $\Omega_{\ell, i}$ top semantically related topics are utilised in equation \ref{eq:semantic_rel}. The most performant value and the next best value are highlighted in \textbf{bold} and \emph{italic} faces respectively.} \label{tab:sr_topics}
\centering \small
\begin{tabular}{l c c c}
  \hline
  Number of Topics ( $\Omega_{\ell, i}$) & Prec. & Rec. & F1 \\
  \hline
  Most Related Topic & \textbf{0.7717} & 0.9431 & \textit{0.8359} \\
  3 Most Related Topics & 0.7622 & \textit{0.9486} & 0.8325 \\
  5 Most Related Topics &0.7659 & \textbf{0.9490} & 0.8345 \\
  10 Most Related Topics & 0.7654 & \textbf{0.9490} & 0.8342 \\
  All Related Topics & \textit{0.7714} & 0.9467 & \textbf{0.8370} \\
  \hline
\end{tabular}
\end{table}

\begin{table}[ht!]\centering \small
\caption{Predictive performance of Semantic TrueLearn model (our proposal) using Precision (Prec.), Recall (Rec.) and F1 Score (F1). The most performant value is highlighted in \textbf{bold} face. The Semantic TrueLearn Model that outperform baseline model ($p< 0.01$ in a one-tailed paired t-test) are marked with $\cdot^{(*)}$.} \label{tab:st_model}
\begin{tabular}{l l l l}
\hline
 & \multicolumn{1}{c}{Prec.} & \multicolumn{1}{c}{Rec.} & \multicolumn{1}{c}{F1} \\
\hline
TrueLearn Novel & \textbf{0.5829} & 0.7924 & 0.6471 \\
\emph{Semantic} TrueLearn & 0.5711 & \textbf{0.8563}$^{(*)}$ & \textbf{0.6512}$^{(*)}$ \\
\hline
\end{tabular}
\end{table}

\subsection{Discussion}
It is evident from Table \ref{tab:results} that incorporating semantic relatedness leads to improvements in overall F1 score in majority of the SR metrics beating the baseline TrueLearn algorithm. Four Semantic TrueLearn models (ones that use \texttt{M\&W}, \texttt{W2V}, \texttt{PMI} and \texttt{BA} 
) tend to outperform the baseline TrueLearn Novel model in terms of precision and F1. The remainder seem to lead in recall.
Entity embedding-based SR metric (\texttt{W2V}) leads to the best performing model. This is expected as neural-based semantic relatedness measures often outperform their graph-based counterparts \cite{ponza_semantic_relate}. Our empirical results in Table \ref{tab:sr_topics} also showed that using \emph{all} semantically related topics gives best results contrary to restricting to some of them. Finally, Table \ref{tab:st_model} shows the superiority of Semantic TrueLearn in comparison to the baseline model that does not exploit SR information from Wikipedia. This is a clear indication that a knowledge base such as Wikipedia can be critical to improving the assumptions used for a learner, modelled using a probabilistic graphical models in the education context. Semantic relatedness can be truly valuable in early stages of the user session when the interaction data about the user is limited, thus addressing the cold-start problem.  

The 
correlation evaluation presented in Figure \ref{fig:performance} (left) shows the lack of correlation between Positive Label Rate and recall score across both TrueLearn models. Although it has been demonstrated by the original authors that TrueLearn algorithm capitalises on recall, there is no information in the work regarding the positive label rate in the datasets. This observation confirms that TrueLearn family of algorithms find true patterns in learner data rather than merely capitalising on the positive labels to boost performance.   

Multiple observations in Figure \ref{fig:performance} (left) give evidence of the superiority of Semantic TrueLearn exploiting the semantic relatedness between topics to boost recall. The main two observations are the new model's stronger Spearman's rank correlation with learner \emph{Avg. Connectedness} and \emph{Min. Cut Set Size}. This is a strong indication that the Semantic TruLearn model is exploiting the topic correlations. The correlation between the number of events, number of unique topics and topic connectedness cause the higher correlation between these features and Semantic TrueLearn model. The Figure \ref{fig:performance} (right) clearly shows how the recall score of predictions is much larger in Semantic TrueLearn algorithm regardless in early or latter stage of the learner session. Linking this to results in Table \ref{tab:results} shows that this impressive gain of recall score is achieved with a much smaller sacrifice of precision score.

\paragraph{Limitations} Amid the significant gains, we observe that most KCs encountered by the model in a session are highly correlated to each other. This leads to overlapping information being propagated repeatedly when using equation \ref{eq:semantic_rel} which may lead to overestimation of knowledge of unseen KCs. Equation \ref{eq:semantic_rel} only acknowledges the existence of correlations between seen and unseen topics (dotted lines in Figure \ref{fig:semantic_rel_problem}) not the correlation within seen topics (solid lines in Figure \ref{fig:semantic_rel_problem}) which is an issue. As the proposed method primarily infers \emph{unobserved} skills, its use diminishes over time when the user session matures (as new topics are encountered less often). Mechanisms to keep using semantic awareness to refine representations is a much needed improvement to the proposed method.   


\section{Conclusions} \label{sec:conclusion}

Leveraging semantic relatedness between Wikipedia topics has demonstrated promise to improving the predictive performance of informational recommenders such as TrueLearn that are built on Wikipedia ontology and probabilistic graphical models.
In addition, we identify that restricting the number of related topics 
leads to degraded performance, suggesting the use of all available knowledge components extracted from Wikification. Our analysis also shows that topic connectedness within learner sessions is positively correlated with the performance gains of Semantic TrueLearn, giving clearer evidence of the positive impact of incorporating this aspect when modelling learners and their journey within an education setting. 


The proposed model is a stepping stone to accounting for semantic relatedness. However, it still disregards the correlation among the observed topics. To address this, 
we propose the following research avenues:
1) using algorithms such as PageRank \cite{pagerank} to derive uncorrelated skill parameters, 
2) accounting for inter-skill correlation in equation \ref{eq:gauss} and 3) building a hierarchical representation of knowledge \cite{pelanek2020managing} consisting of mutually exclusive (uncorrelated) Wikipedia concepts. 
It may also be fruitful to consider richer ontologies \cite{auer2007dbpedia} that contain more fine-grained relationships, entity definitions/categorisations and constrains in the place of raw Wikipedia graph to incorporate finer grain semantic-awareness to the learner model. Mechanisms to continuously utilise SR information 
should be identified and investigated in future work.

Moreover, semantic relatedness measures are not usually built and validated with educational datasets or topics, which is a limitation. In the future, we also aim to  validate the SR metrics with education and informational recommender focused datasets.
As Semantic TrueLearn builds a sub-symbolic representation that is humanly-intuitive, it is possible to create narratives and intelligent user interfaces (e.g. \cite{x5learn,perez2021x5learn}) used to interpret and rationalise \cite{riedl2013interactive} the learnings of the model leading towards more \emph{human-in-the-loop} artificial intelligence that allows verification and scrutinisation of the models \cite{filip_explainable}.   

\section{Acknowledgments}

This  research  is  conducted  as  part  of  the  X5GON  project (\url{www.x5gon.org}) funded from the EU’s Horizon 2020 research and innovation programme grant No 761758.  
We gratefully acknowledge support and funding from the U.S. Army Research Laboratory and the U. S. Army Research Office, and by the U.K. Ministry of Defence and the U.K. Engineering and Physical Sciences Research Council (EPSRC) under grant number EP/R013616/1. This work is also partially supported by the European Commission funded project "Humane AI: Toward AI Systems That Augment and Empower Humans by Understanding Us, our Society and the World Around Us" (grant 820437) and the EPSRC Fellowship titled "Task Based Information Retrieval" (grant EP/P024289/1).

\bibliography{main2}

\end{document}